\documentclass[a4paper,fleqn]{cas-dc}

\usepackage[sort,numbers]{natbib}

\begin{document}
\let\WriteBookmarks\relax
\def\floatpagepagefraction{1}
\def\textpagefraction{.001}
\shorttitle{The efficiency of moderating neutron detector:
  Monte Carlo simulation, experimental validation, and angular dependence }
\shortauthors{C.R. Brune}

\title [mode = title]{The efficiency of moderating neutron detector:
  Monte Carlo simulation, experimental validation, and angular dependence }
\author{Carl R. Brune}[orcid=0000-0003-3696-3311]
\ead{brune@ohio.edu}
\affiliation{organization={Department of Physics and Astronomy and %
                           Institute for Nuclear and Particle Physics},
             addressline={Ohio University}, 
             city={Athens},
             state={Ohio 45701},
             country={United States of America}}

\begin{abstract}
This paper describes the Monte Carlo simulation and experimental
validation of the neutron detection efficiency in a $4\pi$ polyethylene
moderating detector. The experimental validation was performed using
several sources providing neutron energies from 10s of keV up to 14~MeV.
The Monte Carlo simulations employed the {\sc mcnp6} code are believed
to be accurate within 3\% for neutron energies below 6~MeV.
A new method for parametrizing the dependence of the efficiency on
the neutron angular distribution for in-beam experiments is also presented.
This approach utilizes a Legendre decomposition of the efficiency
and allows all of the needed information to be extracted from a single
simulation.
\end{abstract}


\maketitle

\section{Introduction}

For the detection of neutrons on the MeV scale, a detection system
based on thermal neutron detectors embedded in a hydrogenous moderator
is attractive for several reasons.
Such detectors can be very efficient, with
typical total efficiencies exceeding 10\% for neutron sources
located in the interior of the detector. At the same time, they are
extremely {\em insensitive} $\gamma$-rays, which makes these
detectors ideal for low-background applications.
Finally, due to the excellent moderating capabilities of hydrogenous
materials such as polyethylene,
these detectors are relatively compact, with the overall linear
dimensions being under one meter.
I also note that gas-filled proportional counter tubes,
containing either ${}^3{\rm He}$ for BF$_3$, are the thermal neutron
detector of choice.
The reactions ${}^3{\rm He}(n,p)$ and ${}^{10}{\rm B}(n,\alpha)$
have large thermal cross sections and positive $Q$~values, providing
a very clean signal of thermal neutron detection.

Moderating detectors well suited for in-beam measurements of neutrons, which
is the focus of the present work. They have seen wide application
for the measurement of neutron production cross sections for nuclear
astrophysics and other applications~\cite{Cse21,Li22,Bra22}.
The measurement of very small $(\alpha,n)$ cross sections for nuclear
astrophysics is particularly noteworthy use of these detectors.

The first detector of this type was described by Hanson and McKibben
in 1947~\cite{Han47}. It utilized the ``long counter'' geometry, where
a single thermal neutron detector tube is surrounded by moderator and the
neutron source is located outside of the detector and on its symmetry axis.
A subsequent foundational contribution was the 1.5-m-diameter
graphite sphere described by Macklin in 1957~\cite{Mac57}.
It utilized 10 thermal neutron detector tubes embedded in the
spherical moderator and placed the neutron source at the center
of the moderator, thus integrating the neutron yield over emission angles.
Importantly, this work also described how the neutron detection efficiency
depended on the neutron energy and physical properties of the moderator:
size, fast and thermal neutron scattering cross sections and kinematics,
and thermal capture cross section.
Over the following decades, this detector was employed for many
in-beam cross section measurements, e.g., Ref.~\cite{Gib59}.
A detector based on a paraffin moderator was then described by
Marion~{\em et al.} in 1960~\cite{Mar60}. The use of a hydrogenous
moderator instead of graphite significantly reduces the required size of
the detector, but the efficiency becomes more sensitive to neutron energy.
The latter issue was partially mitigated by locating the thermal detector
tubes at two different radii from the beam axis of the detector,
an innovation that is often employed with modern detectors.
The development and use of these detectors continued since this time.
Several additional examples, including the most recent work, are given in
Refs.~\cite{Wes82,Per10,Fal13,Cse21,Li22,Bra22}.

This paper describes the Monte Carlo simulation and experimental validation
of neutron detection efficiency for the polyethylene-moderated detector
described below. This work builds on previous studies in two ways.
The validation is carried out with higher precision
and is extended to 14-MeV neutron energy. More significantly, this
work considers the effect of the reaction angular distribution on the
efficiency. It is only very recently that such effects have been
considered~\cite{Cse21,Li22}. In the present work, a convenient formalism for
quantifying angular distribution effects is presented. With this
method, the Monte Carlo simulation only needs to be performed once, and
then the corrections can be investigated subsequently for various
angular distribution choices.

In the late 1980s and early 1990s, a series of measurements of the
${}^{13}{\rm C}(\alpha,n){}^{16}{\rm O}$ reactions were performed
at Caltech with this detector and published~\cite{Kel89,Bru92,Bru93}.
While these measurements were carefully conducted, the data were analyzed
using oversimplified assumptions about the neutron detection efficiency.
The detection efficiency for ${}^{13}{\rm C}(\alpha,n){}^{16}{\rm O}$
neutrons was assumed to be the same as for ${}^{252}{\rm Cf}$ neutrons.
The mean neutron energy from ${}^{252}{\rm Cf}$ spontaneous fission,
$2.122\pm 0.017$~MeV~\cite{Fro90}, is similar to the energies encountered in
these ${}^{13}{\rm C}(\alpha,n){}^{16}{\rm O}$ measurements.
However the actual energy distributions are quite different.
This issue was revealed by Monte Carlo simulations and experiments
performed a few years later~\cite{Wre98,Bru99,Wre00}.
In the present work, I therefore consider the
${}^{13}{\rm C}(\alpha,n){}^{16}{\rm O}$ reaction and the energy
range of the aforementioned measurements: $0.45 \le E_\alpha \le 1.6$~MeV.
The simulations reported here would thus be a step towards
correcting the previous measurements for an incorrect detection efficiency.

\section{Description of the detector}
\label{sec:polycube}

The $4\pi$ moderating neutron detector is described in
Refs.~\cite{Bru92,Bru93,Wre98,Bru99,Wre00}, with Wrean~\cite{Wre98,Wre00}
providing the most detailed information.
Briefly, the detector consisted of a polyethylene cube of moderator
40~cm on each side. Twelve ${}^3{\rm He}$-filled proportional counters
were arranged in approximate cylindrical symmetry at a radius of
$\approx 12$~cm from the beam axis.
A 11.0~cm~$\times$~11.5~cm open channel
along the beam axis allowed the insertion of the beam pipe and target chamber.
A solid graphite moderator was used to fill the open channel from 1~cm
beyond the end of the target chamber to the outside of the cube.
More details are provided by Figs.~6.1, 6.2, and 6.5 of Ref.~\cite{Wre98}.
For the majority of the measurements, signals from one of the counters
were not processed due intermittent noise issues with that counter.
This counter at the bottom of the counter ring, just to the left of the
center line, as depicted in Fig.~1 of Ref.~\cite{Wre00}.
During the course of the experiments, measurements with a
${}^{252}{\rm Cf}$ source located very near the target position were
frequently utilized to monitor the detector performance and efficiency.

\section{Simulation and validation of neutron detection efficiency}
\label{sec:simulation}

The determination of the neutron detection efficiency for a moderating
detector, such as the one described above, is a non-trivial task.
The efficiency, defined in this work to be the probability of detection,
depends on both the energy and direction of emission of the neutron.
For a detector using a hydrogenous moderator, the dependence on neutron
energy and direction tends to be enhanced because of the large thermal
neutron absorption cross section of natural hydrogen: the efficiency depends
on where in the detector the neutrons become thermalized, which in turn
depends on the neutron emission energy and direction.
Placing the neutron detection counters at a range of distances from the source
may be used to minimize the energy sensitivity and/or to learn about the
distribution of neutron energies~\cite{Mar60,Wes82,Per10,Fal13,Bra22},
although this may come at the cost of detection efficiency.
Most detectors, including the one described here, have an approximate axial
symmetry whose axis matches the beam direction, provided the detector is used
to measure neutrons from reaction experiments.
In addition, in-beam experiments require an opening in the detector to
accommodate the beam pipe and target chamber. As a consequence of these
detector design features, the detector is necessarily not spherically
symmetric and the efficiency must
have some dependence upon the neutron emission direction.

\subsection{Simulation}

\begin{figure}
\includegraphics[width=\columnwidth]{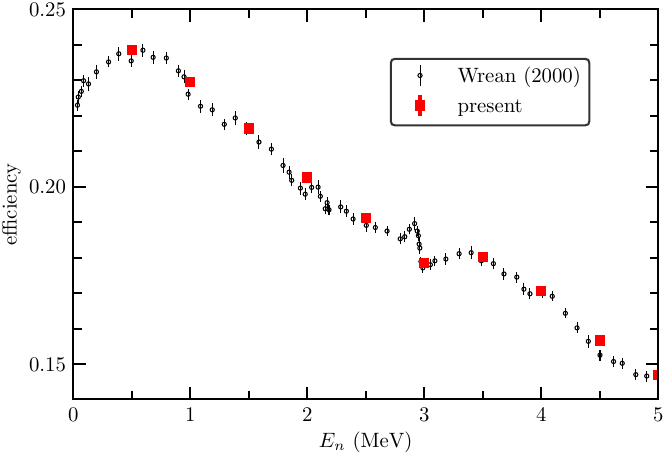}
\caption{The simulated efficiency for isotropic neutrons, calculated
with 12 active counters. The open circles are the results of
Ref.~\cite{Wre00} and the solid red squares are the present simulations.
The error bars on the circles represent statistical error; the statistical
errors in the present simulations are smaller than the size of the points.}
\label{fig:eff_wrean}
\end{figure}

The detector has been simulated using the Monte Carlo program
{\sc mcnp6}~\cite{Goo12}. For the ${}^{13}{\rm C}(\alpha,n){}^{16}{\rm O}$
reaction in the energy range under consideration, reaction
kinematics uniquely define the neutron energy as a function of the beam energy
$E_\alpha$ and the laboratory neutron emission angle $\theta_n$, since
excited states of ${}^{16}{\rm O}$ are not accessible.
Since the target is fairly light, the neutron energy dependence with angle
is considerable. For example, at $E_\alpha=1$~MeV, the neutron energy
varies from 3.2~MeV at $\theta_n=0^\circ$ to 2.4~MeV at $180^\circ$.
The angular distribution may also deviate substantially from isotropy,
due to angular distribution effects in the center-of-mass system as
well as the conversion to the laboratory system.
All of these effects are easily incorporated into the simulation, but requires
the center-of-mass angular distribution be specified.

The simulations reported here are a re-implementation of the model described
in Refs.~\cite{Wre98,Bru99,Wre00}. The new simulation reproduces all of
the key benchmarks described in Refs.~\cite{Wre98,Wre00}, including the
efficiency for an isoptropic source shown in Fig.~3 of Ref.~\cite{Wre00}.
The comparison, for a limited energy range, is shown in
Fig.~\ref{fig:eff_wrean}. The structures in the curve arise
from resonances in ${}^{12}{\rm C}+n$ scattering.
The simulations utilized neutron cross sections based on the
ENDF/B-VII.1 evaluation~\cite{Cha11}, with the {\tt poly.20t} and
{\tt grph.20t} $S(\alpha,\beta)$ libraries being used for thermal
neutrons in polyethylene and graphite, respectively.
The use of these $S(\alpha,\beta)$ libraries was found to be essential
for obtaining simulations that agree with experiment.
Uncertainties in the input cross sections and $S(\alpha,\beta)$ libraries
do not contribute appreciably to the error budget, compared to other
sources of uncertainty.
Other important inputs include the dimensions and densities of the detector.
The density of polyethylene, graphite, and the ${}^3{\rm He}$ gas
are particularly important in the present case.

\begin{figure}
\includegraphics[width=\columnwidth]{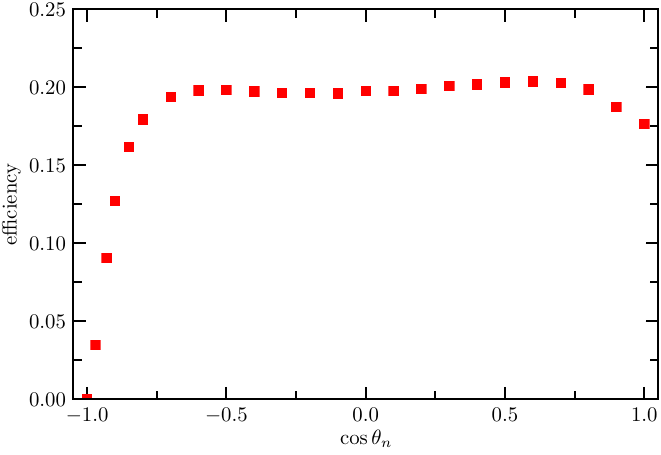}
\caption{The simulated efficiency versus the cosine of the neutron emission
angle, where the angle is defined relative to the beam direction.
The calculation is for a neutron energy of 2~MeV and 11 active counters.
The statistical errors are smaller than the size of the points.}
\label{fig:eff_angle}
\end{figure}

Efficiency curves for an isotropic source are not directly
applicable for the analysis of reaction data, but they do provide
a useful overview of the efficiency.
For reactions measured with unpolarized beams and targets, the neutrons are
distributed uniformly in the azimuthal angle around the beam direction.
I thus always average over the azimuthal angle in the simulations.
An example of the variation of the efficiency with the neutron emission angle
for $E_n=2.0$~MeV is shown in Fig.~\ref{fig:eff_angle}.
Additional information regarding the angle dependence is available in
Ref.~\cite[Fig.~6.7]{Wre98}.
For the specific radioactive source or reaction experiments, the details
of the neutron source were included in the simulation.
This would include the precise source location in the detector,
the source housing, target backing material, and/or beam spot on
target distribution.

\subsection{Validation}

\begin{table}[tbh]
\caption{Comparison of measured and simulated efficiencies for three
  different neutron sources~\cite{Wre98,Wre00}.}
\label{tab:validation}
\begin{tabular}{clll} \hline
  source & \multicolumn{1}{c}{$\langle E_n\rangle$} &
  \multicolumn{1}{c}{experiment} &
  \multicolumn{1}{c}{\sc{mcnp6}} \\ \hline
${}^{252}{\rm Cf}$  & \phantom{1}2.12 & 0.196(6) & 0.1927(15) \\
AmBe               & \phantom{1}4.46 & 0.155(6) & 0.1500(14) \\
${}^3{\rm H}(d,n)$ & 14.1 & 0.0468(14) & 0.0481(2) \\ \hline
\end{tabular}
\end{table}

As discussed in Refs.~\cite{Wre98,Wre00}, the simulation has been
experimentally validated in four ways.
Three of these are comparisons of efficiency measurement to simulation,
as shown in Table~\ref{tab:validation}, which provides the results
from Refs.~\cite{Wre98,Wre00}.
The fourth validation is the energy dependence of the near-threshold
${}^7{\rm Li}(p,n)$ neutron yield. These validations now discussed.

The first efficiency validation used a ${}^{252}{\rm Cf}$ spontaneous
fission source that was calibrated with 3\% accuracy, with 11 of
the detector tubes analyzed.
This source is modeled using the
Watt spectrum parameters given by \citet{Fro90}, which imply
an average neutron energy of $\langle E_n\rangle=2.12$~MeV.
The present simulation yields an efficiency of 0.1919 with negligible
statistical uncertainty for this ${}^{252}{\rm Cf}$ source,
in excellent agreement with the previous simulation and measurement.

The second efficiency validation is used a ${}^{241}{\rm Am}$-Be source
calibrated to 3.5\% accuracy.
This source was modeled in Refs.~\cite{Wre98,Wre00} assuming the
spectrum given by \citet{Gei75}, which has $\langle E_n\rangle=4.46$~MeV.
As shown in Table~\ref{tab:validation}, this simulated value is in
good agreement with source measurement.
The simulation has been repeated using the ${}^{241}{\rm Am}$-Be spectrum
measured by \citet{Klu82}. This spectrum is softer,
having $\langle E_n\rangle=4.19$~MeV, and is consistent with later measurements
and evaluations~\cite{Tho17}. This results in a 3.4\% increase in the
simulated efficiency, which improves the agreement with the calibrated source,
although both simulations agree with the measurement within uncertainties.

The third efficiency validation utilized neutrons from the
${}^3{\rm H}(d,n){}^4{\rm He}$ reaction produced using a thin
Ti-${}^3{\rm H}$ target and a mean deuteron energy of 96~keV.
The $\alpha$-particle yields were measured
separately in a scattering chamber using the same target and deuteron energy.
Because this reaction is isotropic the center-of-mass system, this procedure
calibrates neutron yield per incident deuteron, independent of the
${}^3{\rm H}$ areal density.
This source produces neutrons with $\langle E_n\rangle=14.1$~MeV.
The impact of center-of-mass motion on the simulation was investigated
and found to be negligible.
The ${}^3{\rm H}(d,n){}^4{\rm He}$ comparison is for
six counters: every other counter going around the ring was removed from
detector and disconnected from the electronics, approximately reducing
the efficiency by a factor of two.
As shown in Table~\ref{tab:validation}, good agreement was found between
experiment and simulation for this measurement.

The final validation was a measurement of the ${}^7{\rm Li}(p,n)$
reaction near threshold, where the neutron laboratory angular distribution
and energy change rapidly with beam energy.
As shown and discussed in Refs.~\cite{Wre98,Wre00}, the simulated
efficiency varies by $\approx 10\%$ in the near-threshold region.
Here, the energy dependence of the efficiency-corrected neutron yield agrees
very well with that reported by \citet{Gib59}, with the average deviation
being 1.6\%.

Following Refs.~\cite{Wre98,Wre00}, the uncertainty in the simulated
efficiency is taken to be 3\% for neutron energies below 6~MeV, assuming
that the angular distribution of the reaction is known.

\section{Simulation of neutrons generated from a nuclear reaction with a
  two-body final state}

I now provide more details about the case where the neutrons are generated
by a beam of incident energy $E_0$ bombarding a thin target,
such that beam energy loss can be neglected.
The distribution of neutrons in energy and direction is
then determined by $E_0$, the neutron angular distribution in the
center-of-mass system $W(\theta)$, and kinematics.
The angular distribution may be expressed in the standard manner using
normalized Legendre coefficients
\begin{equation}
W(\theta) = 1 + \sum_{k=1}^{k_{\rm max}} a_k \, P_k(\cos\theta),
\end{equation}
where $P_k$ are the Legendre Polynomials and
$\theta$ is the center-of-mass angle.
I have also introduced the maximum Legendre order $k_{\rm max}$.
It is then logical and straightforward to formulate a Monte Carlo
simulation based on this physics: events are generated in the center-of-mass
system according to $W(\theta)$ and are then transforming to the lab system.
This procedure is implemented in {\sc mcnp6} using the rejection method
for $W(\theta)$ and relativistic kinematics in a custom source subroutine.
If the Monte Carlo simulation uses $N_{\rm sam}$ samples and $N_{\rm det}$
neutrons are detected, the Monte Carlo estimate for the efficiency is given
simply by
\begin{equation}
\langle\varepsilon\rangle_{\rm MC} = \frac{N_{\rm det}}{N_{\rm sam}}.
\end{equation}

The above approach requires performing simulations for every combination of
$E_0$ and $W(\theta)$ needed for the analysis.
This becomes less convenient
if one needs to perform an energy convolution near a narrow resonance where
the $a_k$ may vary rapidly with energy, or if one wants to investigate
different choices or uncertainty quantification for the $a_k$.
I describe below a more convenient and useful approach.
The efficiency can be considered as a function of $\theta$, and being a
well-behaved function, can be decomposed into
Legendre components $\varepsilon_k$:
\begin{equation} \label{eq:legendre_comp}
\varepsilon(\theta)=\sum_{k=0}^\infty (2k+1) \,
  \varepsilon_k \, P_k(\cos\theta) .
\end{equation}
The efficiency averaged over angles is then
\begin{subequations}
\begin{align}
\langle\varepsilon\rangle &= \frac{1}{2}\int_{-1}^1 \varepsilon(\theta) \,
  W(\theta) \, d(\cos\theta) \\
  &=   \varepsilon_0 +\sum_{k=1}^{k_{\rm max}} a_k \,\varepsilon_k ,
  \label{eq:eff_w}
\end{align}
\end{subequations}
keeping in mind that all quantities also depend on $E_0$.
The quantity $\varepsilon_0$ is efficiency for an angular distribution that
is isotropic in the center-of-mass system.
Equation~\ref{eq:legendre_comp} may be inverted to yield
\begin{equation} \label{eq:eps_k_int}
\varepsilon_k = \frac{1}{2}\int_{-1}^1 \varepsilon(\theta) \,
  P_k(\cos\theta) \, d(\cos\theta) .
\end{equation}
If the $\varepsilon_k$ coefficients are known, then using
Eq.~(\ref{eq:eff_w}), the efficiency $\langle\varepsilon\rangle$
can be easily calculated for {\em any} angular distribution.

Equation~(\ref{eq:eps_k_int}) provides the basis for extracting all of
the needed $\varepsilon_k$ coefficients from a single Monte Carlo
simulation performed with $W(\theta)=1$.
This equation may be re-cast as
\begin{equation} \label{eq:eps_k_prob_int}
\varepsilon_k = \varepsilon_0 \int_{-1}^1 \rho(x) \, P_k(x) \,dx ,
\end{equation}
where $x=\cos\theta$ and
\begin{equation}
\rho(x) = \frac{\varepsilon(x)}{2\varepsilon_0} .
\end{equation}
The function $\rho(x)$ may now be interpreted as a probability
distribution, as it is non-negative and normalized:
\begin{equation}
\int_{-1}^1 \rho(x) \, dx = 1 .
\end{equation}
Now suppose there are $N_{\rm det}$ samples of $x$ from the distribution
$\rho(x)$. The Monte Carlo estimate of $\varepsilon_k$ based on
Eq.~(\ref{eq:eps_k_prob_int}) is then~\cite[Subsec.~42.5, p.~750]{Nav24}
\begin{equation}
(\varepsilon_k)_{\rm MC} = \frac{\varepsilon_0}{N_{\rm det}}
  \sum_{i=1}^{N_{\rm det}} P_k(x_i) .
\end{equation}
The final step is to realize that the distribution in $x=\cos\theta$
of the detected neutrons for an center-of-mass isotropic source is proportional
to $\varepsilon(\cos\theta)$ and is hence given by $\rho(x)$.
One can thus use the detected neutron events to provide the samples $x_i$.
Noting that $\varepsilon_0=N_{\rm det}/N_{\rm sam}$, one can also write
\begin{equation} \label{eq:eps_k_mc}
\varepsilon_k = \frac{1}{N_{\rm sam}} \sum_{i=1}^{N_{\rm det}} P_k(x_i) ,
\end{equation}
which is valid for all $k$ and where the MC subscript has been dropped.
This procedure could be generalized by choosing the events from a
distribution $W(x)$, which would be a form of importance sampling.
In this case, the result for the efficiency coefficients would be
\begin{equation}
\varepsilon_k = \frac{1}{N_{\rm sam}} \sum_{i=1}^{N_{\rm det}}
  \frac{P_k(x_i)}{W(x_i)}.
\end{equation}
However, for detectors whose efficiency has little $\theta$ dependence there
is correspondingly little advantage to using this approach.

It is in this manner that the $\varepsilon_k$ can be extracted from a single
Monte Carlo simulation with a center-of-mass isotropic angular distribution.
In {\sc mcnp6}, a custom tally subroutine is used to evaluate
Eq.~(\ref{eq:eps_k_mc}), which depends on the center-of-mass neutron
emission angle.
The efficiency for any angular distribution can now be calculated
using Eq.~(\ref{eq:eff_w}).
Note that, for $k>0$, one expects $|\epsilon_k| \ll \epsilon_0$,
to the extent the detector is insensitive to angular distributions.
It is also expected that $|\varepsilon_k|$ will continue to decrease
as $k$ increases, due to the relatively weak dependence upon angle of the
detector geometry and spatial averaging imposed by the neutron moderation
process.

\begin{figure}
\includegraphics[width=\columnwidth]{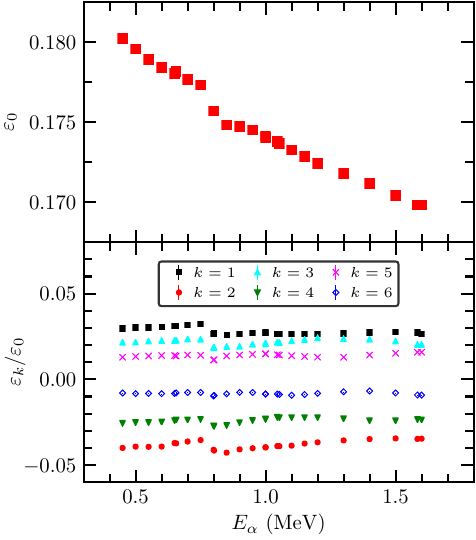}
\caption{The simulated detection efficiency for the
${}^{13}{\rm C}(\alpha,n){}^{16}{\rm O}$ reaction versus $E_\alpha$,
for 11 tubes being analyzed.
The top panel shows the efficiency for center-of-mass isotropy and
the lower panel shows the higher Legendre coefficients, scaled by
$\varepsilon_0$.
The statistical uncertainties are smaller than the size of the points.}
\label{fig:eff_k}
\end{figure}

The simulated $\varepsilon_k$ for the ${}^{13}{\rm C}(\alpha,n){}^{16}{\rm O}$
reaction are shown in Fig.~\ref{fig:eff_k}, for
$0.4\le E_\alpha \le 1.6$~MeV, the energy range relevant for
Refs.~\cite{Kel89,Bru92,Bru93}.
The efficiency for center-of-mass isotropy, $\varepsilon_0$,
is observed to decrease with increasing $E_\alpha$.
For $k>0$, the signs of $\varepsilon_k$ vary according to $(-1)^{k+1}$ and
$|\varepsilon_k/\varepsilon_0|$ is small
compared to unity and decreasing in magnitude with increasing $k$.
The structures arising from ${}^{12}{\rm C}+n$ resonances are smoothed
out, compared to Fig.~\ref{fig:eff_wrean}, but some structure near
$E_\alpha=0.8$~MeV remains, due to the $E_n\approx 3$~MeV
${}^{12}{\rm C}+n$ resonance.
The simulation indicates that actual efficiency for
${}^{13}{\rm C}(\alpha,n){}^{16}{\rm O}$ is 10-30\% lower
than assumed in the original publications~\cite{Kel89,Bru92,Bru93},
with the exact correction depending upon $E_\alpha$ and the angular
distribution.
Although the mean neutron energy from the ${}^{252}{\rm Cf}$ source and
the ${}^{13}{\rm C}(\alpha,n){}^{16}{\rm O}$ reaction for
$0.4\le E_\alpha\le 1.6$~MeV are similar, the energy
distributions are quite different.
The neutrons from ${}^{252}{\rm Cf}$ fission have an approximate
Maxwellian distribution~\cite{Fro90}, peaking at $E_n \approx 0.7$~MeV and
including and exponential tail. Considering Fig.~\ref{fig:eff_wrean}, this
results in an efficiency considerably higher than for
${}^{13}{\rm C}(\alpha,n){}^{16}{\rm O}$, where $2.1 \le E_n \le 3.8$~MeV
for the range of $E_\alpha$ under consideration.

\section{Discussion}

The efficiency coefficients $\varepsilon_k$ are seen to vary slowly
and smoothly with beam energy, with the minor excursions being caused
by ${}^{12}{\rm C}+n$ resonances.
On the other hand, the angular distribution coefficients
$a_k$ may vary rapidly with energy, depending on the resonance structure
of the reaction under study. The use of the Monte Carlo formalism
presented here decouples the assumed angular distribution
coefficients from the simulation.
The simulations thus only need to be performed once for a given beam
energy. It is also sensible to calculate them for an energy grid and
then utilize interpolation.

It is often the case that the angular distribution is not known.
The question then arises: What should one assume?
This formalism allows various assumptions to be investigated.
In this context it is also useful to consider the physics involved.
The odd $a_k$ arise from interference between resonances with
different spins and parities and are may be positive or negative
with roughly equal probability.
While interference effects are also present in the even $a_k$, these
coefficients largely arise from intrinsic angular momentum coupling of the
resonance. In practice, these angular distributions are overwhelmingly
found to have positive $a_k$ coefficients, which implies forward and
backward peaking of the angular distribution.

I now consider a reaction proceeding through a resonance of well-defined
total angular momentum $J$ and parity, and ignore interference effects.
The transition matrix element connection orbital angular
momentum $\ell$ and channel spin $s$ in the initial state to
orbital angular momentum $\ell'$ and channel spin $s'$ in the
final state is given by $T^J_{s'\ell',s\ell}$.
The contribution of $s\ell\rightarrow s'\ell'$ to the angular distribution
coefficient is then~\cite{Bla52,Lan58}
\begin{equation} \label{eq:w_racah}
\begin{split}
a_k \propto ~ &(-1)^{s-s'} (2\ell+1)(\ell\ell 00|k0)W(\ell J\ell J;sk) \\
  &\times (2\ell'+1)(\ell'\ell' 00|k0) W(\ell'J\ell'J;s'k) \\
  &\times |T^J_{s'\ell',s\ell}|^2 ,
\end{split}
\end{equation}
where the $(|)$ symbols are Clebsch-Gordan coefficients and the $W$s are
Racah coefficients. These $a_k$ coefficients
are often quite simple.
For the case of the ${}^{13}{\rm C}(\alpha,n_0){}^{16}{\rm O}$ reaction,
one has $s=s'=1/2$ and for a particular $J^\pi$ only a single $\ell$-$\ell'$
combination is possible.
The resulting coefficients are shown in Table~\ref{tab:w}, where, for
$J\le 7/2$, all of the coefficients are seen to be positive.
It is also noteworthy that they of order unity, implying that
they may give rise to significant efficiency effects.
To summarize, the even $a_k$ coefficients are nearly always positive.
However, these must be investigated for each case, as there are
exceptions. An extreme counter example is the
${}^{12}{\rm C}({}^{12}{\rm C},\alpha){}^{20}{\rm Ne}^*$ reaction,
which only proceeds through $J^\pi=0^+,2^+,4^+\ldots$ states in the
compound nucleus. When populating un-natural parity states of
${}^{20}{\rm Ne}$, the differential cross section for this reaction
vanishes identically at $\theta=0^\circ$ and~$180^\circ$~\cite[p.~311]{Vog68},
which can be confirmed with Eq.~(\ref{eq:w_racah}).

\begin{table}[tbh]
\caption{Non-zero isolated resonance angular distribution coefficients for
${}^{13}{\rm C}(\alpha,n_0){}^{16}{\rm O}$.}
\label{tab:w}
\begin{tabular}{ccccc} \hline
$J$ & $a_0$ & $a_2$ & $a_4$ & $a_6$ \\ \hline
1/2 & 1 \\
3/2 & 1 & 1 \\
5/2 & 1 & 8/7 & 6/7 \\
7/2 & 1 & 25/21 & 81/77 & 25/33 \\ \hline
\end{tabular}
\end{table}

In the statistical or Hauser-Feshbach limit, the results are very
similar. The odd $a_k$ are assumed to average to zero, with the
cross section being described by transmission coefficients which
represent an average over many resonances.
The resulting angular distribution formula is essentially the
same~\cite[Eqs.~(130-132)]{Kon23}, but with additional Racah coefficients
present to convert transmission coefficients from the $jj$ coupling
basis to the channel spin basis.
Again, angular distributions calculated using the Hauser-Feshbach
formalism are found to have overwhelmingly positive even $a_k$ coefficients,
implying forward and backward peaking.
Semiclassical reasons for this finding are given in Refs.~\cite{Eri58}
and~\cite[p.~311]{Vog68}.
The discussion above also suggests that using Hauser-Feshbach $a_k$
coefficients is a better approximation than assuming that
they are zero, even in situations where the statistical approximation
is not necessarily valid.

If the reaction can populate multiple final states in the residual
nucleus, there will be a set of $\varepsilon_k$ coefficients for
each final state. In order to proceed with the analysis, the final-state
branching ratio must also be known.

This methodology has been applied in a recent study of the
${}^{13}{\rm C}(\alpha,n)$ reaction over the energy range
$2.9 \le E_\alpha \le 8.0$~MeV~\cite{Bra23}, although that work provided
little explanation regarding the method of the analysis.
The supplemental information for that paper does provide the assumed
final-state branching ratios as well as the $\varepsilon_k$ and $a_k$ assumed
for each final state branch.
These data could thus be easily reanalyzed if additional information
becomes available about the final-state branchings or angular distributions.

\section{Conclusions}

The results of Monte Carlo simulations of a $4\pi$ polyethylene-moderated
neutron detector have been presented. The simulation has been validated
experimentally for neutron energies of 10s of keV to 14 MeV.
No rescaling of the simulation was needed to achieve excellent agreement
with the measurements.
The uncertainty in the simulated efficiency is estimated to be 3\% for
neutron energies below 6~MeV, assuming the energy and directional
distributions of the neutrons are known.
This level of precision represents a significant step forward, compared
to previous work, which reported uncertainties of 5\%~\cite{Per10},
5\%~\cite{Fal13}, 8\% or 9\% (depending on configuration)~\cite{Cse21},
4\%~\cite{Li22}, and 16\%~\cite{Bra22}, respectively.
All of these works also applied significant renormalizations to their
simulations in order to bring them in to agreement with experiment.
This is also the first instance of a validation being performed up to 14-MeV
neutron energy that I am aware of.

A new method for analyzing the neutron efficiency for in-beam reaction
measurements is also presented. In this framework, the dependence of
the efficiency on the angular distribution is simply parametrized in
terms of Legendre coefficients. This approach allows all of the needed
information to be extracted from a single Monte Carlo simulation, as
opposed to performing multiple simulations for each angular distribution
under consideration. It also provides a straightforward approach
corrected experimental results, if new information about the angular
distribution becomes available.

\section*{Acknowledgments}

I thank James deBoer, Michael Jeswald, and Patricia Wrean for useful
discussions related to this work.
This work was supported in part by the U.S. Department of Energy,
under Grants No.~DE-FG02-88ER40387 and No.~DE-NA0004247.

\bibliographystyle{elsarticle-num-names-nourl.bst}
\bibliography{polycube}

\end{document}